\documentclass[a4paper]{article}
\usepackage{epsfig,amsmath,amsfonts,amssymb,setspace,multirow,textcomp}
\usepackage[T1]{fontenc}
\makeatletter
\renewcommand{\@evenfoot}{\hfil \thepage \hfil}
\renewcommand{\@oddfoot}{\hfil \thepage \hfil}
\makeatother

\renewenvironment{thebibliography}[1]{\begin{oldthebibliography}{#1}\setlength{\parskip}{0ex}\setlength{\itemsep}{0ex}}{\end{oldthebibliography}}

\begin{document}
\fontsize{11}{11}\selectfont 
\title{Detection of C2, CN and CH radicals in the spectrum of the transiting hot Jupiter 
HAT-P-1b}
\author{B.E.~Zhilyaev, M.\,V.~Andreev, S.\,N.~Pokhvala, I.\,A.~Verlyuk}
%
\date{\vspace*{-6ex}}
\maketitle
\begin{center} {\small $Main Astronomical Observatory, NAS \,\, of Ukraine, Zabalotnoho \,27, 03680, Kyiv, Ukraine$}\\
{\tt zhilyaev@mao.kiev.ua}
\end{center}

\begin{abstract}
In this paper we report spectroscopy of the transiting hot Jupiter HAT-P-1b. The HAT-P-1b is a giant ($R = 1.2 RJ$), low-mean density transiting extrasolar planet in a visual binary system, composed of two sun-like stars. The host star HAT-P-1b known as ADS 16402 B is a G0V C dwarf (V = 9.87). We revealed optical emission of $C_{2}$, $CN$ and $CH$ radicals in the spectrum of the hot Jupiter HAT-P-1b. We discovered radial pulsation of the hot Jupiter HAT-P-1b with a period of about 1900 sec.


\end{abstract}

\section*{\sc introduction}
\indent \indent The $C_{2}$ Swan bands were observed in the spectra of some stars \cite{Hobbs} and comets \cite{Churyumov}. The $C_{2}$  in the emission spectrum of a comet are excited through resonance fluorescence with the sunlight \cite{Mayer}. Stockhausen \& Osterbrock \cite{Stockhausen} predicted from calculations with a simple molecule model that the vibrational temperature for cometary $C_{2}$ should be close to the solar color temperature and independent of the comet's heliocentric distance.

Spectroscopy of hot Jupiter HAT-P-1b during transits with the grating spectrograph STIS with R = $\lambda / \Delta \lambda$ = 500 aboard the Hubble Space Telescope revealed strong optical absorbers (such as the Na I doublet at $\lambda$ = 5893 $\AA$ \cite{Nikolov}. The authors speculated that the best fit for the average dayside temperature of HAT-P-1b is 1500 $\pm$  100 K. Optical emission of $C_{2}$ and other radicals were not detected. 

Ground-based optical spectroscopy of hot Jupiter HAT-P-1b during transits had been observed with the Gemini Multi-Object Spectrograph (GMOS) instrument on the Gemini North telescope \cite{Todorov}. The authors found that the resulting transit spectrum is consistent with previous Hubble Space Telescope observations. However, the authors do not detect the Na resonance absorption line at $\lambda$ = 5893 $\AA$. Their 520 - 950 nm observations reach a precision comparable to that of HST transit spectra in a similar wavelength range. However,  GMOS transit between 320 - 800 nm suffers from strong systematic effects and yields larger uncertainties.

Transmission spectroscopy has proven a powerful method to study the atmospheres of transiting exoplanets. This technique uses the differing wavelength-dependence of absorption and scattering processes in planetary atmospheres, resulting in a wavelength-dependent planetary radius \cite{Alexader}.

Our goal is to study the emission spectrum of the planet. Emission spectra are observed in the atmospheres of comets excited by solar radiation.

An atom or molecule in open space can be detected by means of resonant absorption and reemission of electromagnetic waves, known as resonance fluorescence, which is a fundamental phenomenon of quantum optics. In three-dimensional (3D) space, experimentally achieved extinction can reach 12\% of transmitted power \cite{Astafiev}, \cite{Gerhardt}, \cite{Muller}.

In this work, optical emission of the $ C_ {2} $, $ CN $ and $ CH $ radicals in the spectrum of the hot Jupiter HAT-P-1b was discovered. Section 2 describes a series of spectra obtained with a low-resolution spectrograph. Our study of hot Jupiter HAT-P-1b during transit focused on the structure of the $C_{2}$ bands. In the following sections, we describe the analysis techniques used. The observed sequence $C_{2}$ $\Delta \nu=-1$ was used in Section 3 to estimate the excitation temperature. The features of the spectra and the main results are discussed in Section 4.

\section*{\sc OBSERVATIONS AND DATA PROCESSING}

The hot-Jupiter HAT-P-1b is one of the first known exoplanet. Planetary parameters are \cite{Nikolov}: Period  $P$ = 4.46529976 $\pm$  (55) days. A semi-major axis 0.05561 AU. Mass $MP = 0.525 MJ$. Radius $RP = 1.319 RJ$. Density $\rho \sim 0.282\, g \,cm^{-3}$. Equilibrium temperature $T_{eq}$ 1322 $\pm$ 15 K. Incident flux $F = 0.699 \cdot 10^{9}\, erg\, s^{-1}\, cm^{-2}$. Total transit duration 0.11875 days.

HAT-P-1b was observed at the Terskol Observatory  (North Caucasus, 3100 m at sea level) with the Carl Zeiss 0.6-meter telescope with an imaging slitless spectrograph \cite{Zhilyaev}. The spectrograph provides a spectral resolution of R $\simeq $ 150 for $10^{m}$ stars with a moderate signal-to-noise ratio. A series of 900 measurements were obtained with a temporal resolution of 6.6 s (Fig. 1).

The start of exposure was on 27.11.2009, 17:22:16 UT. The observed minimum was at 17:48:40 UT. The calculated minimum from ephemeris is at JD = 2455163.24239 and corresponds to 17:49 UT on 27.11.2009. O - C = 20 sec.

\begin{figure}[!h]
\centering
\begin{minipage}[t]{.45\linewidth}
\centering
\epsfig{file = 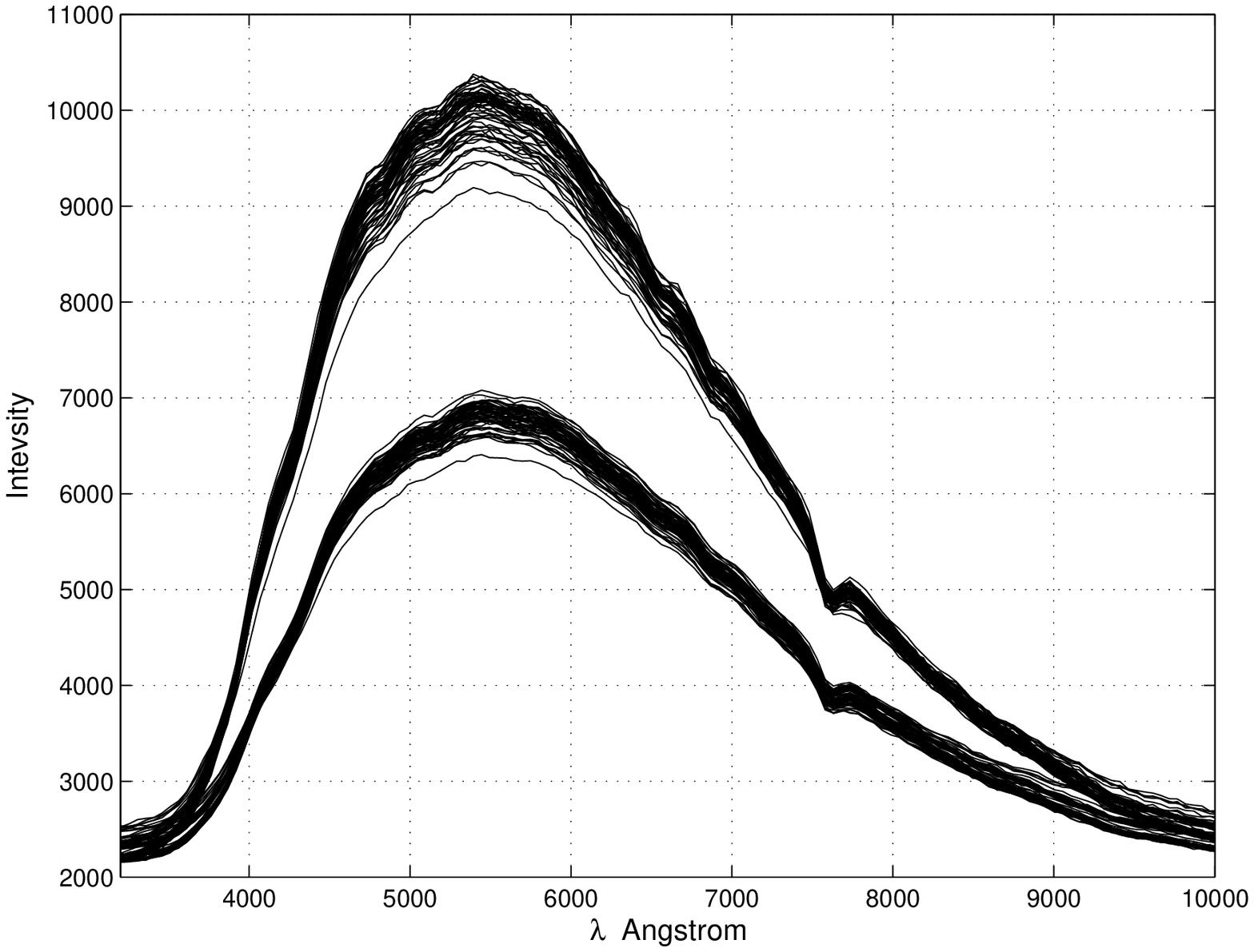,width = 1.0\linewidth} \caption{The spectra summarized up to an integration time of 99 sec.}\label{fig1}
\end{minipage}
\hfill
\begin{minipage}[t]{.45\linewidth}
\centering
\epsfig{file = 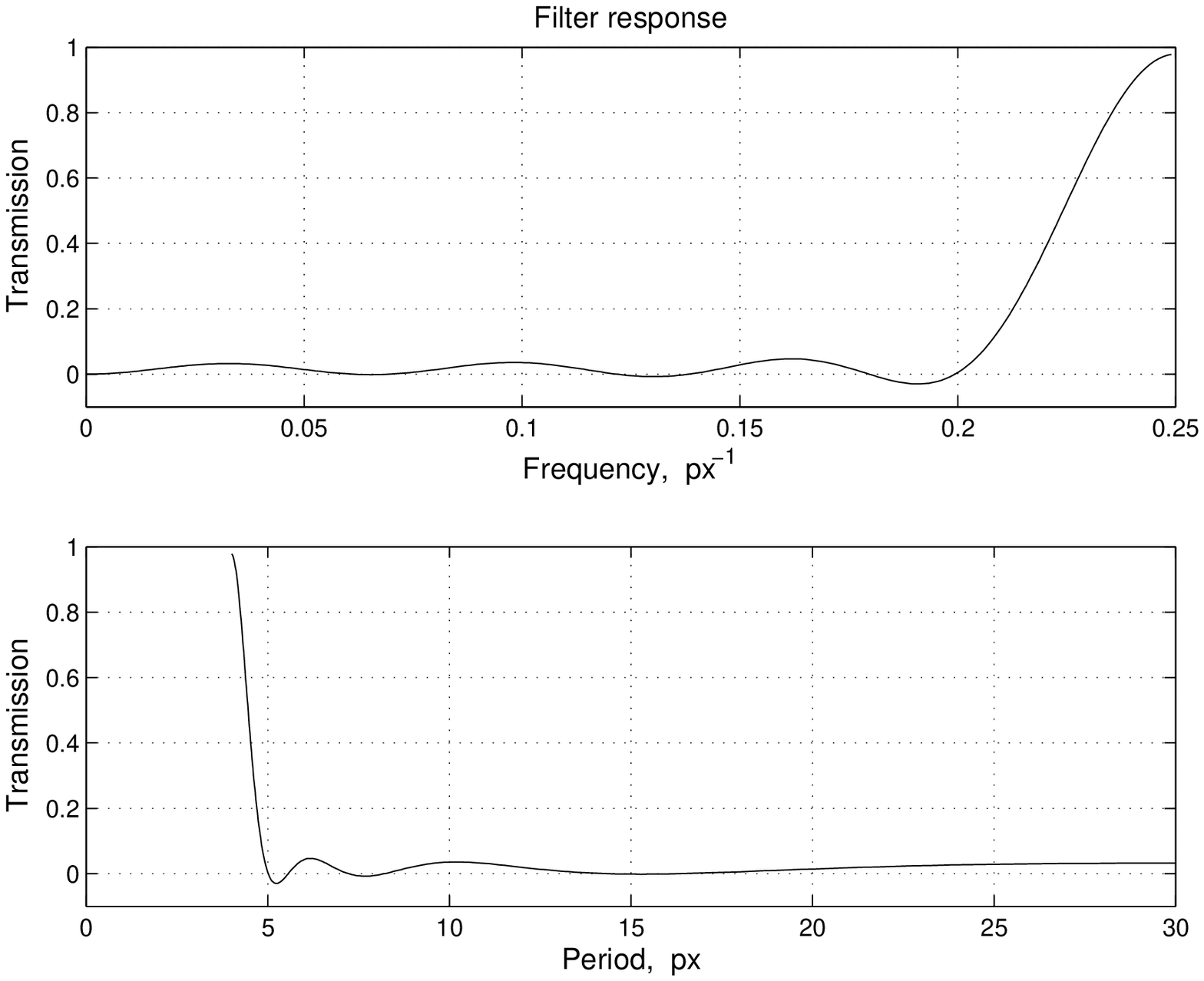,width = 1.0\linewidth} \caption{The filter transmission curves in px$^{-1}$ and px scale.}\label{fig2}
\end{minipage}
\end{figure}


\section*{\sc Diagnostics of radical emissions}

The spectra of the flame of carbon compounds were studied in Johnson \cite{Johnson}. It was concluded that Swan bands are clearly visible in the flame of burning hydrocarbons. The author gives detail of all the structures at a wavelength of the head of the $\lambda \lambda$4382, 4737, 5165, 5635, 6191 $\AA$ groups.

Our task is to find own variability in the spectral lines, using the difference in the noise spectrum and the spectrum of the useful signal. The solution of the problem is carried out in stages:

\begin{figure}[!h]
\centering
\begin{minipage}[t]{.45\linewidth}
\centering
\epsfig{file = 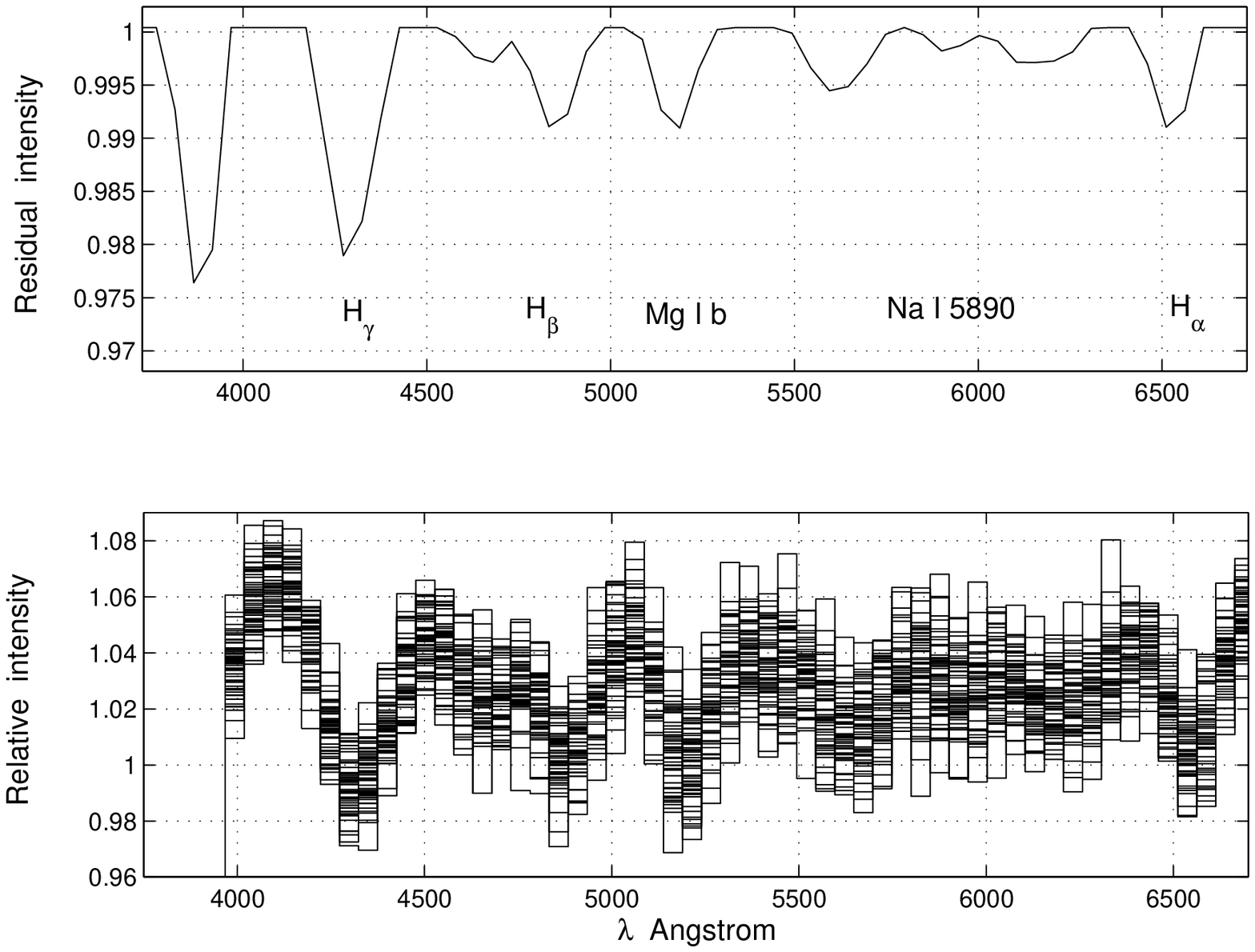,width = 1.0\linewidth} \caption{The absorption spectrum of HAT-P-1b (top figure). The emission spectra of HAT-P-1b (bottom figure).}\label{fig3}
\end{minipage}
\hfill
\begin{minipage}[t]{.45\linewidth}
\centering
\epsfig{file = 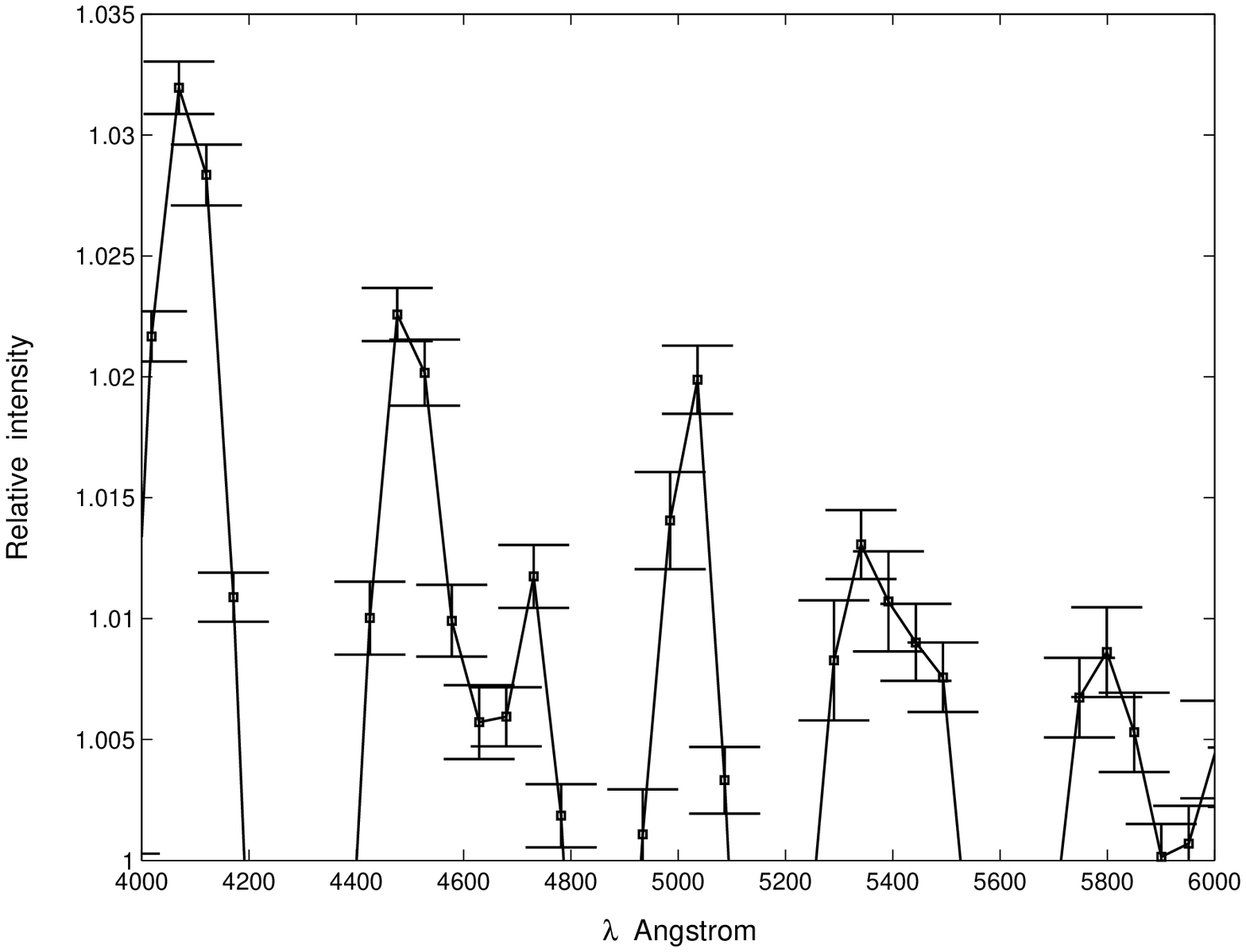,width = 1.0\linewidth} \caption{The averaged emission spectrum of HAT-P-1b.}\label{fig4}
\end{minipage}
\end{figure}

\begin{itemize}   
\item Elimination of interference caused by changes in atmospheric transparency using the automatic gain control method.

\item Alignment of data using the cross-correlation method to eliminate shifts caused by
instrument bending and atmospheric refraction.

\item Calculation of the spectra map, the matrix whose rows are wavelengths, and the
columns are the power spectra of the intensity variations at each spectrum wavelength.
\end{itemize} 

Detecting the emission component in the spectrum is complicated, especially if its intensity is comparable to that of noise. The task is somewhat simplified for broadband emission.
The spectra have been averaged to detect very weak emission components.  A special fitting procedure based on the algorithm developed by the authors was used.

To eliminate the absorption spectrum, the high pass FIR filter was used. We use a Kaiser window to design the filter with pass band frequency 0.2 px$^{-1}$, pass band ripple 0.5 dB, and stop band attenuation of 26 dB \cite{Kaiser}. 
The filter transmission curves are shown in Figure 2. The spatial frequency is in pixels, the pixel is equal to 3 Angstroms in the wavelength scale.

The digital filtering method makes it possible to find a weak emission component in the spectrum. A significant amount of data (900 spectra) allowed us to detect an emission of several percent of the intensity of the continuum.

Figure 3 demonstrates the operation of the algorithm on the entire array of HAT-P-1b spectra. The top figure saw the low-resolution absorption spectrum of HAT-P-1b. The Balmer hydrogen lines H$_{\alpha}$, H$_{\beta}$, H$_{\gamma}$ are clearly seen, as well as the Na resonance absorption line at $\lambda $ = 5893 $\AA$. The bottom figure saw the emission in the spectrum of HAT-P-1b obtained as a result of filtration. It is seen that the emission peaks significantly exceed the measurement errors.

Figure 4 demonstrates the averaged emission spectrum of HAT-P-1b with the measurement errors.

Even a short examination of the emission flux distribution reveals several spectral features of HAT-P-1b (Fig. 3, 4). Three features due to $C_{2}$ are seen, corresponding to the Swan-band sequences with $\Delta \nu=1$ ($\lambda $4425 - $\lambda $4629), $C_{2}$ $\lambda $5450 - $\lambda $5650 ($\Delta \nu=-1$); $C_{2}$ $\lambda $5700 - $\lambda $5800 ($\Delta \nu=-1$). Three features of $CN$ are seen also, corresponding to  $\lambda $3458, $\lambda $3510, and $\lambda $9258 (1,0).

\begin{figure}[!h]
\centering
\begin{minipage}[t]{.45\linewidth}
\centering
\epsfig{file = 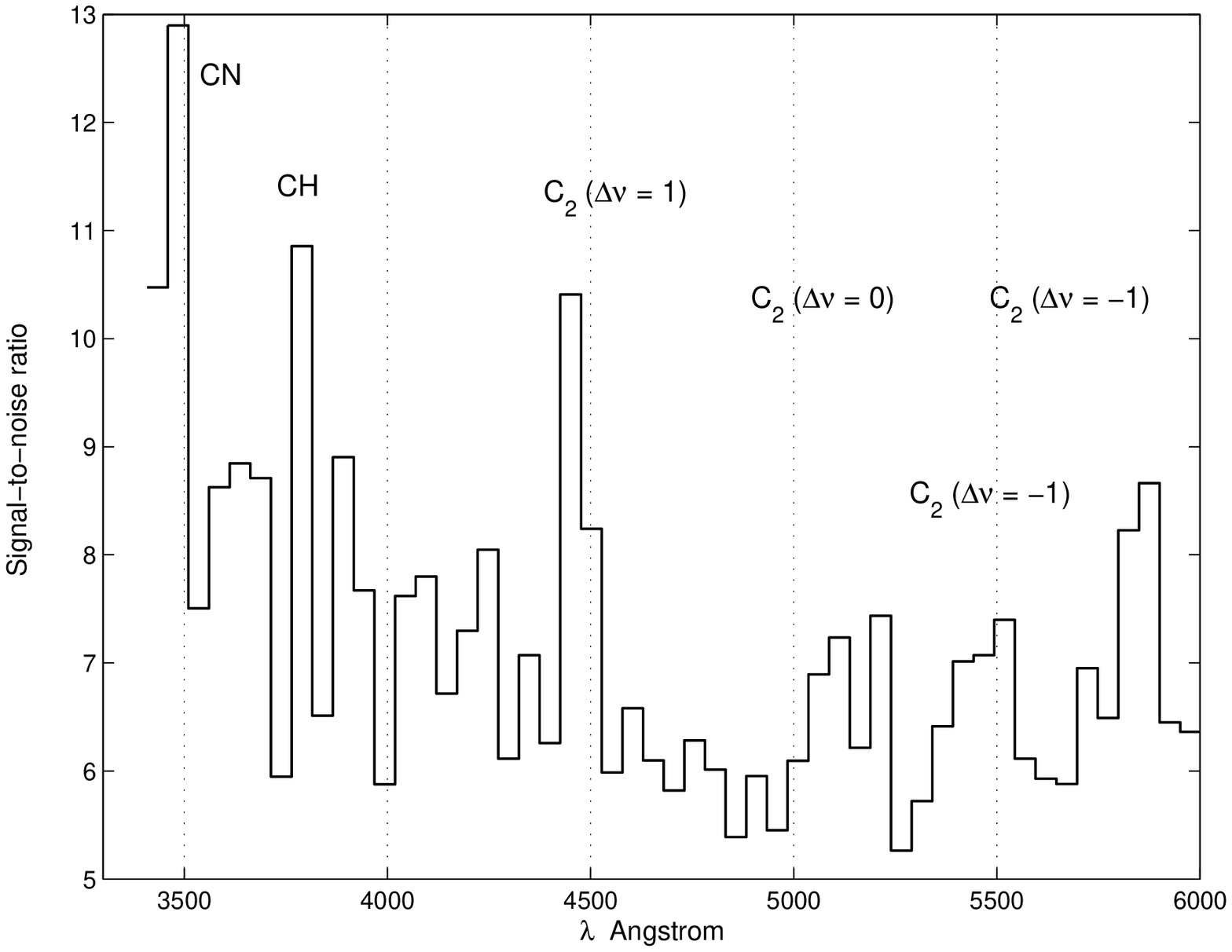,width = 1.0\linewidth} \caption{Variations in the emission spectrum of HAT-P-1b.}\label{fig5}
\end{minipage}
\hfill
\begin{minipage}[t]{.45\linewidth}
\centering
\epsfig{file = 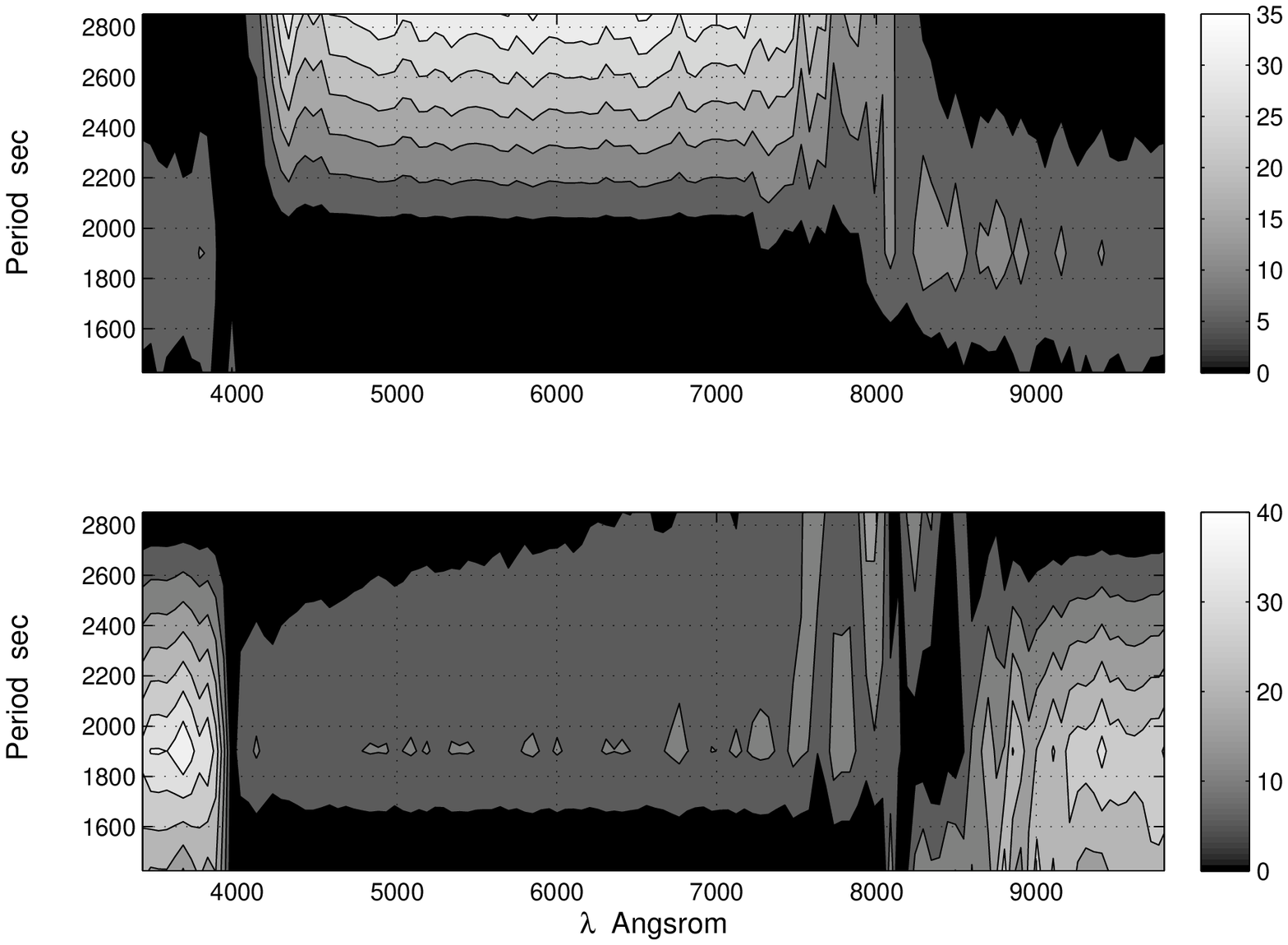,width = 1.0\linewidth} \caption{The SNR map spectra of HAT-P-1a (top picture) and HAT-P-1b (bottom picture).}\label{fig6}
\end{minipage}
\end{figure}

Fullerton \cite{Fullerton} developed a method for detecting line variations, called "The Temporal Variance Spectrum" (TVS). This method compares statistically the deviations in the spectral lines with the deviations in nearby regions of the continuum. If the deviations in the spectral line are larger than the deviations  in the continuum, then it can be argued that a variation is detected at a certain level of statistical significance.

The TVS technique with a signal-to-noise ratio (SNR) in spectra allows detection of variations in lines.  

The intensities of the pulsed flows are shown in Fig. 5. HAT-P-1b exhibit variations at the wavelengths of the $\lambda \lambda$4382, 5165, 5635 $\AA$ $ C_ {2} $ groups, $ C_ {2} $ 5700 - 5800 $\AA$.

The picture also demonstrates the activity of radicals at $\lambda \lambda$3510 and 3800 $\AA$ wavelengths associated with $CN$ ($\Delta \nu=0$) \cite{Brooke} and $CH$ (as Solar $CH$ $\lambda \lambda$3557, 3779 $\AA$) \cite{Hase}.

Figure 6 shows variations in the map spectra of HAP-1b and companion star HAT-P-1a. One can see developed activity in the signal-to-noise ratio scale. Bottom picture reveals the activity in the wavelengths of the $\lambda \lambda$4382, 5165, 5635 $C_{2}$ groups.  The islands of activity (SNR > 10) with a period of about 1900 sec are probably associated with oscillation of the hot Jupiter HATP-1b.

\begin{figure}[h]
\centering
\resizebox{0.50\hsize}{!}{\includegraphics[angle=000]{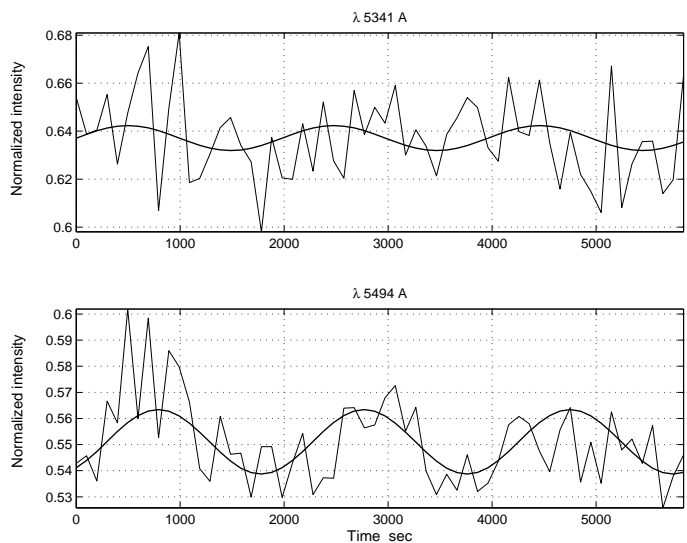}}%
\caption{Normalized light curves of two lines $\lambda \lambda$5341 and 5494 $\AA$.
} \label{figure: HAT_P_1_14qz.eps}
\end{figure}

From Fig. 6 (top picture), we can conclude that the spectrum of the comparison star HAT-P-1a does not demonstrate intensity variations in the wavelengths of the $C_{2}$ groups. Low-frequency variations in the continuum is probably associated with self-variability. It can also be assumed that there is intrinsic variability in the lines of the Balmer series.

\vspace*{1ex}
Figure  8 and Figure 9 allows us to compare our data with a combination of the optical spectra of comet 109P/Swift-Tuttle obtained November 26, 1992, and the spectra of comets Shoemaker-Levy (1991a1) and Hartley-Good (1985).

One can see almost complete coincidence of the spectral features of $C_{2}$ $\Delta \nu=-1$ , wavelength range $\lambda \lambda$5450 - 5650 $\AA$, ($\Delta \nu=-3$), $\lambda$ 6600 $\AA$; $CN$ (2 - 0), (3 - 1) wavelength range $\lambda \lambda$ 7820 - 8039 $\AA$, $CN$ (1 - 0), (2 - 1) wavelength range $\lambda \lambda$ 9109 - 9223 $\AA$.

This similarity practically confirms our conclusions about the detection of radicals in the spectrum of the transiting hot Jupiter HAT-P-1b.

\begin{figure}[h]
\centering
\resizebox{0.50\hsize}{!}{\includegraphics[angle=000]{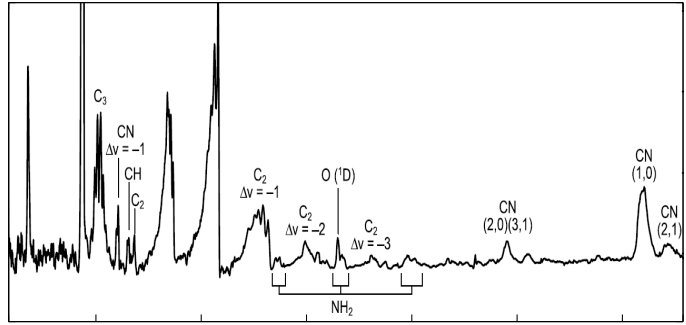}}%
\caption{Combination of the optical spectra of comet 109P/Swift-Tuttle and the spectra of comets Shoemaker-Levy (1991a1) and Hartley-Good (1985). Arbitrary scale.
} \label{figure: HAT_P_1_14qz.eps}
\end{figure}

\begin{figure}[h]
\centering
\resizebox{0.50\hsize}{!}{\includegraphics[angle=000]{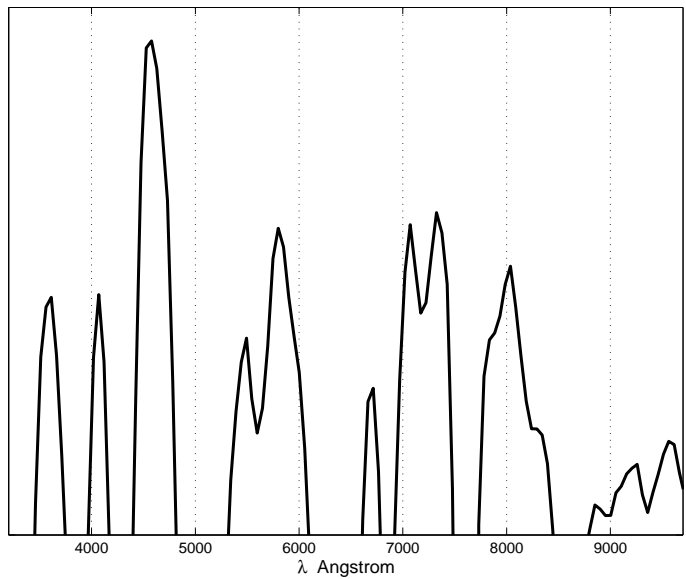}}%
\caption{The emission spectrum of HAT-P-1b. Arbitrary scale.
} \label{figure: HAT_P_1_14qz.eps}
\end{figure}

\vspace*{1ex}
A well-known relationship that determines the pulsation period of stars is \cite{Unno}:

\begin{equation}\label{}
P = Q (\rho_{\odot} / \rho)^{\frac{1}{2}}
\end{equation}
where $P$ denotes the pulsation period, and $\rho_{\odot}$ and $\rho$ denote the mean solar and stellar densities. Essentially, this is free fall time. The theoretical value of $Q$ for the standard model with a polytrope index n = 3 is 0.039 days \cite{Allen}. 
For hot Jupiter HAT-P-1b, the average density is 0.282 g/cm$^{-3}$ \cite{Nikolov}. The pulsation period $P$ is equal to 2122 sec and practically coincides with the observed one.

To determine the excitation temperature $T_{ex}$ we use the estimates of the spectrum intensity in the Swan band. We use the intensity ratio of the $C_{2}$ Swan system $d^{3}\prod_{g}\leftarrow a^{3}\sum_{u} $. Only pure vibrational transitions with $\Delta \nu=-1$ were considered.

We have chosen two lines $\lambda \lambda$5341 and 5494 $\AA$ as successive vibrational levels with intensity ratio 1.19 (Fig. 7, 0.64/0.55). The intensity ratio is then the population ratio for j=0 and j=1:
\begin{equation}\label{}
    \frac{n_{j=1}}{n_{j=0}}=\frac{g_{1}e^{-E_{1}/kT}}{g_{0}e^{-E_{0}/kT}}=3\,e^{-\Delta E/kT}=\frac{I_{1}}{I_{0}}
\end{equation}
where $g_{0}=1$, $g_{1}=3$ are the statistical weights, $E_{i}$ =1.99e-8/$\lambda_{i}$ is the energy of photon \cite{Allen}, $k$ is the Boltzmann constant.

This leads to the temperature
\begin{equation}
    T_{ex}=\Delta E/k/\ln(3/(I_{1}/I_{0})  = 814\, K
\end{equation}

From the intensity ratio of the $C_{2}$ lines with $j$=0 and $j$=1 the excitation temperature of 814 K follows. 


\section*{\sc Conclusions}

Spectroscopy of hot Jupiter HAT-P-1b during transit with the grating spectrograph with R = $\lambda / \Delta \lambda$ $\simeq$ 150 reveals the spectrum of non-resolved Swan bands $C_{2}$ $\lambda \lambda$4425 - 4629 $\AA$ ($\Delta \nu=1$); $C_{2}$ $\lambda \lambda$5000 - 5200 $\AA$ ($\Delta \nu= 0)$; $C_{2}$ $\lambda \lambda$5450 - 5650 $\AA$ ($\Delta \nu=-1$); $C_{2}$ $\lambda \lambda$5700 - 5800 $\AA$ ($\Delta \nu=-1$); $CN$ $\lambda \lambda$3448 - 3509 $\AA$ and $CH$ $\lambda$ 3779 $\AA$.

The TVS technique with a signal-to-noise ratio (SNR) in spectra allowed to detect variations in lines of radicals.  

Calculation of the spectra map, that are the power spectra of the intensity variations at each spectrum wavelength allowed to find the islands of activity with a period of about 1900 sec associated, probably, with oscillation of the hot Jupiter HAT-P-1b.

For the standard model with a polytrope index n = 3 and the average density of 0.282 g cm$^{-3}$, the pulsation period of hot Jupiter HAT-P-1b $P$ was appreciated equal to 2122 sec what practically coincides with the observed within 10\%.

To determine the excitation temperature $T_{ex}$ we use the estimates of the spectrum intensity in the Swan band. We use the intensity ratio of the $C_{2}$ Swan system $d^{3}\prod_{g}\leftarrow a^{3}\sum_{u} $  for sequence $\Delta \nu=-1$.


From the intensity ratio of the $C_{2}$  lines with $j$=0 and $j$=1 a vibrational temperature of 814 K was appreciated. 

We discovered radial pulsation of the hot Jupiter HAT-P-1b with a period of about 1900 sec.


\end{document}